\documentclass[singlecolumn,pra,preprint, showpacs]{revtex4}
\usepackage{amssymb}
\usepackage{stmaryrd}
\usepackage{marvosym}
\usepackage{graphicx}
\usepackage{url}

\newcommand{\unit}[1]{\ensuremath{\,\mathrm{#1}}}
\newcommand{\Og}{\ensuremath{\Omega}}
\newcommand{\Om}{\ensuremath{\Omega_\mathrm{m}}}
\newcommand{\Gm}{\ensuremath{\Gamma_\mathrm{m}}}

\newcommand{\og}{\ensuremath{\omega}}
\newcommand{\meff}{m_\mathrm{eff}}

\newcommand{\tex}{\ensuremath{\tau_\mathrm{ex}}}
\newcommand{\tnul}{\ensuremath{\tau_\mathrm{0}}}
\newcommand{\trt}{\ensuremath{\tau_\mathrm{rt}}}

\newcommand{\fref}[1]{figure \ref{#1}}
\newcommand{\Fref}[1]{Figure \ref{#1}}
\newcommand{\eref}[1]{(\ref{#1})}

\begin{document}

\title[Monitoring micromechanical vibration using whispering gallery mode resonators]%
{High-sensitivity monitoring of micromechanical vibration using
optical whispering gallery mode resonators}
\author{A.~Schliesser{\bf \footnote{These authors contributed equally to this work.}},
 G.~Anetsberger$^{*}$, R.~Rivi{\`e}re, O.~Arcizet and T.~J.~Kippenberg\footnote{Electronic mail: tjk@mpq.mpg.de}}
\affiliation{Max-Planck-Institut f\"ur Quantenoptik,
Hans-Kopfermann-Stra\ss e 1, 85748 Garching, Germany}
%\ead{tjk@mpq.mpg.de}

\begin{abstract}
The inherent coupling of optical and mechanical modes in high
finesse optical microresonators provide a natural, highly sensitive
transduction mechanism for micromechanical vibrations. Using
homodyne and polarization spectroscopy techniques, we achieve
shot-noise limited displacement sensitivities of
$10^{-19}\unit{m}/\sqrt{\unit{Hz}}$. In an unprecedented manner,
this enables the detection and study of a variety of mechanical
modes, which are identified as radial breathing, flexural and
torsional modes using 3-dimensional finite element modelling.
Furthermore, a broadband equivalent displacement noise is measured
and found to agree well with models for thermorefractive noise in
silica dielectric cavities. Implications for ground-state cooling,
displacement sensing and Kerr squeezing are discussed.
\end{abstract}


\maketitle

%\submitto{\NJP}

\section{Introduction}
The transduction and measurement of small displacements of
mechanical oscillators is important in a variety of studies ranging
from macroscale gravitational wave detection \cite{Braginsky1977} to
micron-scale cantilever-based force measurements \cite{Rugar2004}.
One embodiment that is particularly amenable to measurement of small
displacements is the parametric coupling of a high-Q electrical or
optical resonance to a mechanical oscillator \cite{Braginsky1977}.
In the case of a Fabry-Perot interferometer with a harmonically
oscillating end-mirror, this parametric coupling manifests itself as
a position-dependent shift of the resonance frequency, thereby
allowing transducing the mechanical motion into a change of the
phase of a resonant field probing the Fabry-Perot cavity.

However, even for a perfect measurement apparatus, the sensitivity
is limited by the laws of quantum mechanics. Fundamentally, any
linear measurement process entails that it must also exert a
backaction onto the mechanical oscillator, as first discussed by
Braginsky \cite{Braginsky1992,Griffard1976}. Braginksy identified
two types of backaction: \textit{quantum backaction} and
\textit{dynamic backaction}.\ Dynamic backaction occurs when the
optical interferometer (or cavity) is excited in a detuned manner.\
In this case the radiation pressure force exerted by photons in the
cavity can become viscous and lead to either amplification or
cooling of the mechanical motion. This effect is entirely classical
and has been first observed in 2005 in the case of
radiation-pressure amplification and oscillation
\cite{Kippenberg2005, Rokhsari2005} and in 2006 also for
radiation-pressure cooling \cite{Gigan2006, Arcizet2006,
Schliesser2006}. For a recent review see Ref.\
\cite{Kippenberg2007}. For the case of resonant excitation of the
cavity or circuit, this effect can however in principle be entirely
suppressed. In contrast, quantum backaction cannot be suppressed and
arises from the discrete nature of the photons (or electrons)
involved in the measurement process. The quantum fluctuations of the
intracavity field cause a random force that drives the mechanical
oscillator and thereby leads to perturbation of its position. \ This
quantum backaction provides a limit to continuous position
measurements and leads to the so called standard quantum limit
\cite{Braginsky1992, Caves1980}. At the standard quantum limit, the
measurement imprecision is equal to the zero point motion of the
mechanical oscillator, to which both shot noise and quantum
backaction contribute in equal amounts. Over the past decades,
significant progress has been made in approaching this quantum limit
of motion measurement in the context of both electromechanical and
optomechanical experiments. Researchers have investigated a variety
of mechanisms and devices as motion transducers of mesoscopic
oscillators, for example quantum point contacts \cite{Cleland2002,
Poggio2008}, superconducting single-electron transistors
\cite{Knobel2003,LaHaye2004}, atomic point contacts
\cite{Flowers-Jacobs2007}, microwave interferometers
\cite{Braginsky1977, Regal2008} and magnetomotive transducers \cite%
{Ekinci2002, Gaidarzhy2005}. In terms of demonstrated sensitivity,
optical transducers \cite{Hadjar1999, Arcizet2006, Corbitt2007a,
Caniard2007a} have
been unsurpassed, measuring mechanical displacement down to the $10^{-20}%
\mathrm{m}/\sqrt{\mathrm{Hz}}$ level for measurement bandwidths up to $1%
\unit{MHz}$.

Important to the detection process is that the optical transducer is
typically not only sensitive to a single mechanical mode of
interest, but sensitive to any differential change in the cavity's
optical path length. Therefore, a variety of mechanical modes of the
cavity boundary can contribute and the signal from the mode of
interest may be spectroscopically extracted based on its Fourier
frequency. The thermal excitation of other modes, and other
effective cavity length fluctuations may constitute a
measurement background in excess of the quantum shot of the light%
%\cite{Arcizet2006,Caniard2007}
 used to monitor the mode of interest.\ Assessing this background
therefore proves particularly important if one mode is laser-cooled
below the bath temperature as recently achieved \cite{Cohadon1999,
Arcizet2006a, Gigan2006, Kleckner2006, Schliesser2006, Corbitt2007,
Corbitt2007a, Poggio2007, Thomson2007, Schliesser2008}.

In this letter, we provide a broadband analysis of the noises in the
optomechanical transduction of radial displacements in toroidal
silica microcavities, revealing in detail
the influence of all other mechanical modes. Toroidal microcavities \cite%
{Armani2003} are optical resonators that host both high-quality
optical and mechanical modes in one and the same device, and have
been used to demonstrate radiation pressure dynamic backaction for
the first time \cite{Kippenberg2005,Rokhsari2005}. Efficient cooling
by dynamical backaction \cite{Braginsky2002} has been demonstrated
both in the \textquotedblleft Doppler\textquotedblright\ \cite%
{Schliesser2006} and the resolved-sideband regime
\cite{Wilson-Rae2007, Marquardt2007, Schliesser2008}, rendering them
particularly interesting for the goals of the emerging discipline of
cavity quantum optomechanics \cite{Kippenberg2007}, which pertains
to studying quantum phenomena of mesoscopic mechanical oscillators.
We report broadband interferometric measurement of their
radio-frequency mechanical modes based on parametric coupling to the
optical whispering-gallery modes (WGMs). Using adaptations of both
the H{\"a}nsch-Coulliaud polarization spectroscopy \cite{Hansch1980}
and optical homodyne measurement, displacement sensitivities at the
level of $10^{-19}\unit{m}/\sqrt{\mathrm{Hz}}$ are achieved over a
measurement bandwidth of up to $20\unit{MHz}$.

We find sparse spectra of mechanical modes which allow obtaining a
detailed understanding of the modes using 3-dimensional
finite-element simulations.\ More than 20 mechanical modes are
observed between DC$-100$ $\unit{MHz}$ comprising radial breathing,
flexural and torsional modes. We furthermore identify a broadband
noise background which is attributed to thermorefractive noise, as
previously observed in silica microspheres \cite{Gorodetsky2004}.
The detailed understanding of these noise processes (both due to
mechanical modes and thermorefractive noise) is particularly
important for studies such as pondermotive squeezing
\cite{Fabre1994}, ground state cooling \cite{Wilson-Rae2007,
Marquardt2007}, as well as squeezing using the third order Kerr
nonlinearity of glass \cite{Slusher1985}.

\section{Whispering gallery modes for optical motion transduction}
\begin{figure}[htb]
\centering
\includegraphics[width=.6\linewidth]{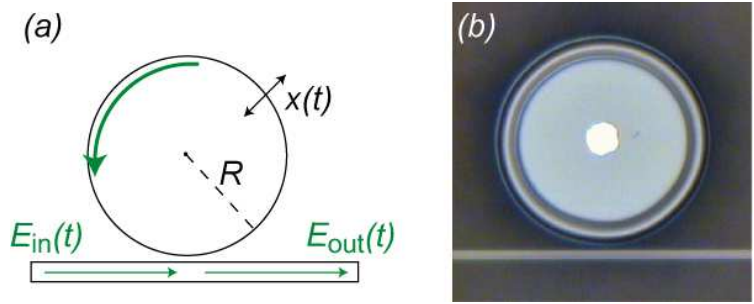}
 \caption{Motion transduction with a whispering gallery mode resonator.
 (a) Changes $x$ is the cavity radius change the resonance condition for
 the field $E_\mathrm{in}$ evanescently coupled to the WGM. Information on the
 displacement $x$ is therefore imprinted on the field $E_\mathrm{out}$ transmitted
 through the taper. (b) Optical microscope image of tapered fibre coupled
 toroidal microresonator (top view).} \label{f:system}
\end{figure}

Decades of research in the field of gravitational wave astronomy
have brought major fundamental \cite{Braginsky1992, Caves1980} and
technological advances in interferometric transduction of mechanical
displacements. More recent efforts \cite{Tittonen1999, Hadjar1999,
Arcizet2006} have shown that these techniques are well suited for
application to much lighter oscillators at the microscale. Such
oscillators are expected to display quantum effects at significantly
higher temperatures. In the following, we will briefly review the
limits for the interferometric detection of micromechanical
oscillations using high-quality WGMs \cite{Ilchenko2006}. The
employed silica toroidal resonators (\fref{f:system}) possess
ultra-high-Q optical modes which are confined by total internal
reflection \cite{Armani2003}. In addition, microcavities also
exhibit structural resonances giving rise to high frequency
vibrational modes \cite{Kippenberg2005,Rokhsari2005,Carmon2007}.
Mechanical modes which affect the circumference of the cavity shift
the optical resonance frequency, and thereby couple to the optical
degree of freedom. This was recognized in early experiments that
demonstrated the parametric oscillation instability
\cite{Kippenberg2005,Rokhsari2005} and provides, as shown here, a
natural way for highly sensitive motion transduction.

A change of the major radius $R$ by a small displacement $x$ induces
a shift of the resonance frequency of the whispering gallery modes
located in the rim of the toroid by an amount $\Delta \og_0/\og_0=x
/R$. This induces a change in the properties of a field launched
into this mode. In the usual coupling geometry using a tapered
fibre, the field transmitted through the tapered fibre reads
\cite{Haus1984}
\begin{equation}
 E_\mathrm{out}(t)=\frac{\tex-\tnul+2 i (\og-\og_0)\tex \tnul}{\tex+\tnul+2 i (\og-\og_0)\tex
 \tnul} \cdot E_\mathrm{in}(t) \label{e:trans}
\end{equation}
where $\tex$ and $\tnul$ are the inverse cavity decay rates due to
output coupling to the taper and due to other losses\footnote{The
times $\tex$ and $\tnul$ can also be expressed in terms of the
probability of transmission $T$ to the taper and loss $L$ to the
environment within the time $\tau_\mathrm{rt}$ of one round-trip via
$\tex=\trt/T$ and $\tnul=\trt/L$.}. The condition with $\tex=\tnul$
is usually referred to as critically coupled or impedance matched.
To first order, the transmitted amplitude is not affected by small
mechanically-induced resonance frequency shifts $\Delta\omega_0$ for
a resonant laser $\og=\og_0$, the phase of the field, however, is.
By comparison with a phase reference in an interferometric
measurement, the displacement $x$ can be thereby be detected. For
example, the output field may be brought to interference with a
strong field $E_\mathrm{lo}$ at the same frequency $\og_0$ using a
half-transmissive beam splitter. Choosing the appropriate phase of
the reference field $E_\mathrm{lo}$, the photon fluxes detected at
the two output ports of the splitter are determined by
 $|(E_\mathrm{out}(t)\pm i E_\mathrm{lo}(t))/\sqrt{2}|^2$.
Subtraction of these two simultaneously measured fluxes yields a
differential signal
\begin{equation}
 h(x)=\frac{8 \omega_0 \eta_\mathrm{c}}{\kappa R}\frac{\sqrt{P_\mathrm{in}
 P_\mathrm{lo}}}{\hbar \og}\cdot x \label{e:slope}
\end{equation}
to first order in $x$, where $\kappa=\tnul^{-1}+\tex^{-1}$ is the
cavity's total decay rate (i.e.\ linewidth) and $P_\mathrm{in}/\hbar
\og$ and $P_\mathrm{lo}/\hbar \og$ are the photon fluxes
corresponding to the fields $E_\mathrm{in}$ and $E_\mathrm{lo}$. The
coupling efficiency
\begin{equation}
\eta_\mathrm{c}=\frac{\tnul}{\tnul+\tex}%
\label{e:eta}
\end{equation}
can take values $0 \ldots 1$, approaching $0$ for undercoupling
($\tex\rightarrow \infty$) and $1$ in the case of overcoupling
($\tex\rightarrow 0$), while at critical coupling it is equal to
$1/2$. Physically this quantity thus describes the probability of an
intracavity photon coupling to the output fibre. The fundamental
noise source in this detection scheme arises from the quantum phase
noise of the light being detected. This leads to a spectral density
of flux (and thus signal) fluctuations of $P_\mathrm{lo}/\hbar
\omega$ in the experimentally desirable limit $P_\mathrm{lo}\gg
P_\mathrm{out}$. The resulting minimum detectable displacement
$\delta x_\mathrm{min}$ is
\begin{equation}
  \frac{\delta x_\mathrm{min}}{\sqrt{\Delta\!f}}=\frac{\tex \kappa^2 R}{8 \og_0
  \sqrt{P_\mathrm{in}/\hbar\omega}}=\frac{\lambda}{16
  \pi \eta_\mathrm{c}\mathcal{F}} \frac{1}{\sqrt{P_\mathrm{in}/\hbar\omega}},
  \label{e:sens}
\end{equation}
where the finesse $\mathcal{F}=c/n R \kappa$ of the WGM was
introduced, $\lambda=2 \pi c/ n\omega_0$ is the optical wavelength
in glass and $\Delta\!f$ the measurement bandwidth.  Note that the
finesse $\mathcal F$ is also affected by the coupling
$\eta_\mathrm{c}$ via
$\mathcal{F}=\mathcal{F}_0(1-\eta_\mathrm{c})$, where
$\mathcal{F}_0$ is the ``intrinsic'' finesse in the undercoupled
limit $\tex\rightarrow\infty$, so that the optimum sensitivity is
achieved at critical coupling $\eta_\mathrm{c}=1/2$.

We note that the same result is obtained following a more formal
approach, using the linearized quantum Langevin equations (QLE) of
the coupled optomechanical system around a stable working point
\cite{Fabre1994,Giovannetti2001,Walls2008}. This allows in
particular calculating the fluctuations in any arbitrary quadrature
of the output field. Comparison of the resulting quantum
fluctuations and mechanically induced fluctuations in the detected
output phase quadrature yields a minimum displacement of
\begin{equation}
  \frac{\delta x_\mathrm{min}(\Omega)}{\sqrt{\Delta\!f}}=\frac{\lambda}{16
  \pi \eta_\mathrm{c}\mathcal{F}}
  \frac{1}{\sqrt{P_\mathrm{in}/\hbar
  \og}}\sqrt{1+\left(\frac{\Omega}{\kappa/2}\right)^2}.\label{e:realsens}
\end{equation}
Compared to \eref{e:sens}, this calculation adds only a correction
due to the finite response time of the cavity for Fourier
frequencies $\Omega$ exceeding the cavity cutoff $\kappa/2$. We note
that the full expression \eref{e:realsens} can also be derived from
a classical calculation of the signal, considering the amplitude of
the motional sidebands of the field coupling out of the cavity and
comparing it with detection shot noise \cite{Schliesser2008}. The
detection limit \eref{e:realsens} corresponds to a spectral density
of measurement imprecision
\begin{equation}
  S_x(\Omega)=\left(\frac{\lambda}{16
  \pi \eta_\mathrm{c}\mathcal{F}}\right)^2\cdot
  \frac{1+\left(\frac{\Omega}{\kappa/2}\right)^2}{P_\mathrm{in}/\hbar\og}.
\end{equation}

As first pointed out by Braginsky \cite{Braginsky1992}, this
measurement inevitably exerts backaction on the mechanical device.
In the case of an optical transducer, quantum backaction is enforced
by the fluctuations of radiation pressure due to a fluctuating
intracavity photon flux. From the linearized Langevin equations, the
intracavity force fluctuation spectrum can be calculated to have the
form
\begin{eqnarray}
 S_F(\Omega)= 16 \eta_\mathrm{c}
 \mathcal{F}^2 \frac{P_\mathrm{in}/\hbar \og
 }{1+\left(\frac{\Omega}{\kappa/2}\right)^2} \hbar^2
 k^2.
\end{eqnarray}
It is noted that $S_x$ and $S_F$ fulfill the expected uncertainty
relation
\begin{equation}
  S_x S_F=\frac{\hbar^2}{4 \eta_\mathrm{c}} \geq
  \frac{\hbar^2}{4}
\end{equation}
with an equality in the limit $(\eta_\mathrm{c}\rightarrow 1)
\Leftrightarrow (\tnul \gg \tex)$, that is, for a strongly
overcoupled cavity. The fluctuating backaction force drives the
mechanical oscillator, and therefore induces position fluctuations.
Assuming the fluctuations in the amplitude and phase quadratures of
the input light field are uncorrelated, the position fluctuations
induced by $S_F$ are uncorrelated with the apparent position
fluctuations $S_x$. Thus the total measurement uncertainty is given
by
\begin{equation}
  S_x^\mathrm{tot}(\Og)=S_x(\Og)+|\chi(\Og)|^2 S_F(\Og)
\end{equation}
where
\begin{equation}
  \chi(\Og)=\frac{1}{\meff(\Om^2-\Og^2-i \Og\Gm)}
\end{equation}
is the susceptibility of the mechanical oscillator and $\Gm$ is its
mechanical damping rate. It is important to note that this
susceptibility is modified when the optical resonance is excited in
a detuned manner \cite{Braginskii1967}. However, for resonant
probing as considered here, it is not modified. The total
measurement uncertainty is minimized for an input flux of
\begin{equation}
  P^\mathrm{in}_\mathrm{opt}/\hbar\og=\eta_\mathrm{c}^{-\frac{3}{2}}\frac{\lambda^2}{128
  \pi^2
  \mathcal{F}^2}\frac{1+\left(\frac{\Og}{\kappa/2}\right)^2}{\hbar  \left|\chi(\Og)\right|}
\end{equation}
yielding
\begin{equation}
 S_\mathrm{x}^\mathrm{SQL}(\Og) =
 \frac{ \hbar\left|\chi(\Og)\right|}{\sqrt{\eta_\mathrm{c}}}=
 \frac{\hbar}{\meff \sqrt{\eta_\mathrm{c} ((\Om^2-\Og^2)^2+\Gm^2\Og^2)}},\label{e:SQL}
\end{equation}
called the {\it standard quantum limit} \cite{Braginsky1992,
Caves1980} in the case $\eta_\mathrm{c}=1$. Its peak value at $\Om$
is
\begin{equation}
S_\mathrm{x}^\mathrm{SQL}(\Om)=\frac{1}{\sqrt{\eta_\mathrm{c}}}\frac{\hbar}{
\meff \Gm \Om}.
\end{equation}

In this calculation we have explicitly considered the effect of the
coupling conditions to the cavity, which can---as a unique
feature---be varied continuously in the experiment by adjusting the
gap between the coupling waveguide and the WGM resonator. The SQL is
approached most closely in the overcoupled limit $\tex\ll\tnul$. It
is noteworthy that the fibre-taper coupling technique to
microtoroids can deeply enter this regime, and $100\cdot \tex<\tnul$
($\eta_\mathrm{c}=99\%$) has been demonstrated \cite{Spillane2003}.
On the other hand, such a strong coupling reduces the cavity finesse
and thus comes at the expense of a higher optimum power
$P^\mathrm{in}_\mathrm{opt}$. Working with weaker coupling, such as
critical coupling as typically pursued in this work, brings only a
moderate penalty as \eref{e:SQL} shows, for example, a factor of
$\sqrt{2}$ for $\eta_\mathrm{c}=1/2$.

In our experiment performed at room temperature, the noise induced
by quantum backaction is masked by thermal noise due to a
fluctuating Langevin force with $S_F^\mathrm{th}(\Og)=\hbar \meff
\Gm |\Og| \coth\left(\hbar |\Og|/2 k_\mathrm{B} T\right)\approx 2
\Gm \meff k_\mathrm{B} T$ as $k_\mathrm{B} T \gg \hbar \Om$, adding
a third term to the total displacement noise
\cite{Fabre1994,Giovannetti2001},
\begin{eqnarray}
   S_x^\mathrm{tot}(\Og)&&=S_x(\Og)+|\chi(\Og)|^2\left(
   S_F(\Og)+ S_F^\mathrm{th}(\Og)\right)\\
   &&=\left(\frac{\lambda}{16
  \pi \eta_\mathrm{c}\mathcal{F}}\right)^2\cdot
  \frac{1+\left(\frac{\Omega}{\kappa/2}\right)^2}{P_\mathrm{in}/\hbar\og}+\nonumber\\
  &&\qquad+ |\chi(\Og)|^2\left(16 \eta_\mathrm{c}
 \mathcal{F}^2 \frac{P_\mathrm{in}/\hbar \og
 }{1+\left(\frac{\Omega}{\kappa/2}\right)^2} \hbar^2
 k^2+2 \Gm
\meff k_\mathrm{B} T\right).
\end{eqnarray}
This  expression constitutes a description of the spectrum that is
measured by analyzing the phase quadrature of the transmitted light
past the microresonator \cite{Giovannetti2001}.

\section{Experimental implementation}
Detection of the phase of a light field with quantum-limited
sensitivity is a standard task in quantum optics, and several
techniques have been developed to achieve this. In the following
section, two techniques which were successfully implemented for
motion transduction in whispering gallery mode microresonators are
described.

\subsection{Homodyne spectroscopy}
\begin{figure}[tb]
\centering
\includegraphics[width=.8\linewidth]{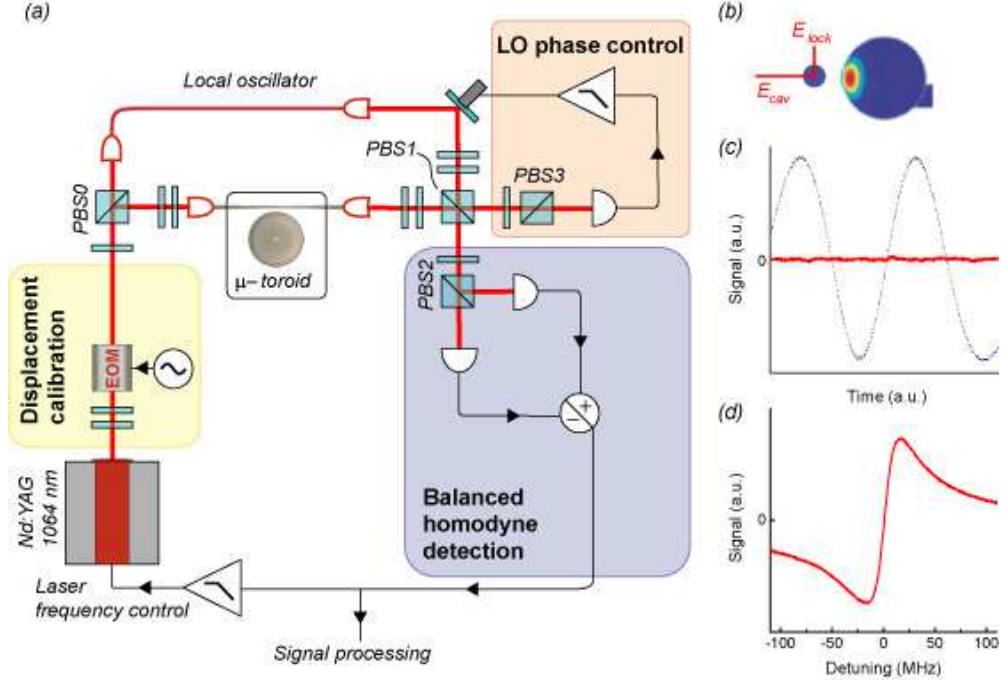}
 \caption{(a) Optical interferometric displacement transducer based on homodyne spectroscopy of
 light transmitted past the cavity (``$\mu$-toroid''). The phase
 of the local oscillator is actively stabilized (``LO phase control''). Details are given
 in the text. PBS0-PBS3, polarizing beam splitters. (b) Cross
 section through the fibre taper and the toroidal rim in the
 coupling region. The polarization in the taper is slightly mismatched with the polarization of the cavity mode.
 Thus only part of the total field $E_\mathrm{cav}$ couples to the WGM, the other component $E_\mathrm{lock}$ can be
 used for the stabilization of the local oscillator phase. The components $E_\mathrm{cav}$ and $E_\mathrm{lock}$
 are separated in PBS1. (c) Signal in the
 balanced receiver for a scanning local oscillator (dotted, blue) at
 low power, and for the locked LO (red). The shown locked trace was
 recorded for about 5 seconds. (d) Typical experimental error signal in the balanced
 receiver when the laser is scanned over a cavity resonance with
 the local oscillator locked to the appropriate phase.
 } \label{f:homodynesetup}
\end{figure}

The most common method for quantum-limited phase measurement is a
balanced homodyne receiver \cite{Yuen1983} as employed in previous
optomechanical experiments \cite{Hadjar1999, Briant2003b,
Caniard2007a}. We briefly discuss the experimental protocol used for
homodyne spectroscopy. This method is adapted to the ring topology
of our resonator by sending the laser beam that is transmitted
through the taper to a beam splitter, where it is brought to
interference with a strong local oscillator (LO), see
\fref{f:homodynesetup}. Signal and local oscillator beams are
derived from a monolithic Nd:YAG laser operating at
$\lambda=1064\unit{nm}$. This source exhibits quantum-limited
amplitude and phase noise at Fourier frequencies $\Omega/2\pi
\gtrsim 5 \unit{MHz}$ and power levels $P_\mathrm{LO}+P \lesssim 5
\unit{mW}$ of interest. Due to its limited tuning speed and range we
use a home-built external-cavity diode laser for
pre-characterization of several samples until a suited toroid is
found. The Nd:YAG laser beam is split using a polarizing beam
splitter (PBS0), the signal beam is sent through the coupling taper
in the near field of the excited cavity mode and its phase is
shifted depending on the mutual laser-cavity detuning. The local
oscillator travels in the reference arm of an effective Mach-Zehnder
interferometer and is recombined with the signal beam at a
polarizing beam splitter (PBS1). Spatial matching of the incident
modes is facilitated by using single-mode fibre as mode filter on
the local oscillator. After spatial recombination, interference is
enforced using a retarder plate and another polarizing beam splitter
(PBS2).

%Note that the vacuum entering at the open port of this combiner does
%not interfere (for a perfect PBS) with the strong LO due to its
%orthogonal polarization at the output and does therefore not degrade
%the signal.

It was found advantageous to actively stabilize the phase of the
local oscillator at the combining beam splitter. Since the phase of
the signal beam depends on the detuning of the laser from the
cavity, which may itself be subject to drifts and fluctuations, it
is not suited as a phase reference. However, by purposely
introducing a small polarization mismatch (cf.\
\fref{f:homodynesetup}(b)) between the light in the taper region and
the either predominantly TE- or TM-like WGM modes of the
microcavity, it is possible to utilize the signal's polarization
component \emph{orthogonal} to the cavity WGM mode as a phase
reference, in order to lock the Mach-Zehnder interferometer to the
desired phase angle between WGM and LO fields. WGM and locking
polarization components in the signal beam are separated by the
first beam splitter (PBS1), after compensation of fibre-induced
polarization rotation. An error signal is created by detecting the
interference between the locking and the LO beam in a third PBS
(PBS3), which provides a feedback signal to the piezoelectric
transducer displacing a mirror.

In order to reach quantum-limited detection sensitivity of the
signal beam's phase, the power of the local oscillator has to be
chosen such that quantum shot noise exceeds the receiver noise,
$P_\mathrm{LO}\gg \mathrm{NEP}(\Og)^2/\eta \hbar \omega$, where
$\mathrm{NEP}(\Og)$ is the noise-equivalent power of the receiver,
and $\eta$ the detection efficiency. The employed commercial InGaAs
receivers provide $\eta(1064\unit{nm})\sim 87\%$, $\mathrm{NEP}\sim
10 \unit{pW}/\sqrt{\mathrm{Hz}}$ between $0$ and $80 \unit{MHz}$.
Balancing the detectors avoids the saturation of the amplifiers by
the large d.c.-field of the local oscillator. Excess losses due to
mode matching and tapered fibre imperfections further reduce the
total detection efficiency to (an unoptimized)
$\eta_\mathrm{tot}\sim 50\%$.

An advantageous feature of the homodyne signal is that its
d.c.-component directly provides a dispersive error signal
\begin{equation}
   h(\Delta\omega)=\frac{2 \eta_\mathrm{c} \kappa
 \, \Delta\og}{\Delta\og^2+(\kappa/2)^2}
\sqrt{P_\mathrm{cav}
   P_\mathrm{LO}}\label{e:homodyne}
\end{equation}
that can be used for locking the laser to the center of the optical
microcavity resonance. At the low signal powers used here, locking
is very stable. An example of an experimentally obtained error
signal is shown in \fref{f:homodynesetup}(d). Simultaneously, the
fluctuations in the differential photocurrent induced by both
optical shot noise in the signal and the thermal noises in the
cavity displacement can be frequency-analyzed using a
high-performance electronic spectrum analyzer. For calibration
purposes, we frequency-modulate the laser using a $\mathrm{LiNbO}_3$
phase modulator external to the laser. The frequency modulation is
given by $\delta \og=\beta \Omega_\mathrm{mod}$ for known modulation
depth $\beta$ and frequency $\Omega_\mathrm{mod}$, and generates the
same signal as would be induced by a radius modulation of $\delta
x=\,R\delta \og/\og$ \cite{Tittonen1999, Hadjar1999, Gorodetsky2004,
Schliesser2008}, \emph{independent} of cavity linewidth and coupling
conditions. If the cavity linewidth is known in addition, the
spectra can be absolutely calibrated at all Fourier frequencies,
taking into account the reduced sensitivity beyond the cavity cutoff
at $\kappa/2$.

\subsection{Polarization spectroscopy}
\begin{figure}[tb]
\centering
\includegraphics[width=.7\linewidth]{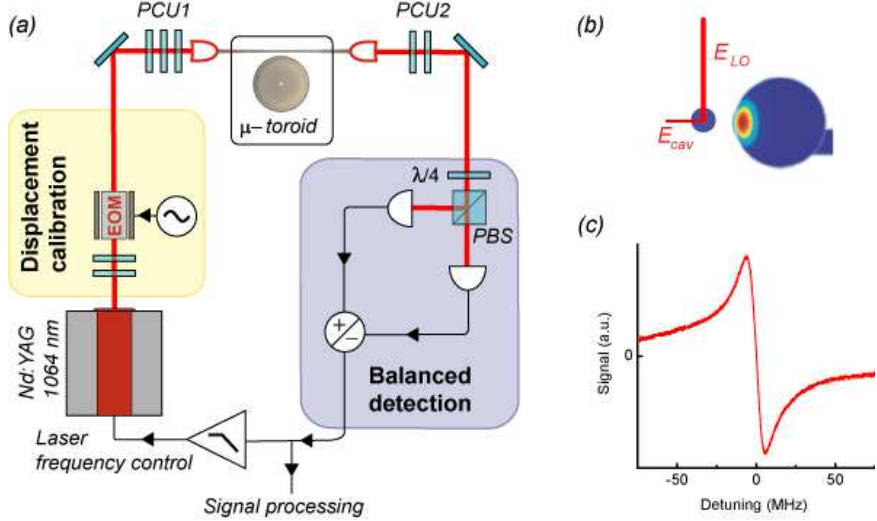}
 \caption{Optical interferometric displacement transducer based on polarization spectroscopy
 of light transmitted in the taper past the cavity (``$\mu$-toroid''). (a) After phase modulation with an
 electro-optic modulator, the polarization is prepared with a first polarization control unit (PCU1).
 The cavity WGM defines signal and LO polarizations. A second
 polarization control unit (PCU2) compensates for fibre birefringence.
 Polarization analysis using a $\lambda/4$ plate and a polarizing
 beam splitter enforces interference of the signal and LO fields.
 (b) Due to the polarization non-degeneracy of the WGM in the
 cavity, only one polarization component of the light interacts with
 the mode. (c) Typical error signal obtained when the laser is
 scanned over a cavity resonance.
  }
 \label{f:haenschcouillaud}
\end{figure}

A simplified setup may be obtained by co-propagating the local
oscillator in the same spatial, but orthogonal polarization mode as
compared to the signal beam \cite{Schliesser2008}. Since the WGM
modes have predominantly TE or TM character and are not degenerate,
this guarantees that the local oscillator is not affected by the
cavity. Due to common-mode rejection of many sources of noise in the
relative phase between signal and LO (for example, frequency noise
in the optical fibre), the passive stability is sufficiently
enhanced to enable operation without active stabilization
(\Fref{f:haenschcouillaud}).

Enforcing interference between local oscillator and signal beams
then corresponds to polarization analysis of the light (comprising
both signal and LO) emerging from the cavity. While novel in the
present context of a tapered fibre coupled microcavity, this is a
well established technique to derive a dispersive error signal from
the light reflected from a Fabry-Perot type reference cavity, named
after their inventors H{\"a}nsch and Couillaud \cite{Hansch1980}.

If fibre birefringence is adequately compensated, the error signal
reads
\begin{equation}
   h(\Delta\omega)=\frac{2 \eta_\mathrm{c} \kappa
 \, \Delta\og}{\Delta\og^2+(\kappa/2)^2}
\sqrt{P_\mathrm{cav}
   P_\mathrm{LO}},
\end{equation}
identical to \eref{e:homodyne},  and a typical trace is shown in
\fref{f:haenschcouillaud}(c). This is used to lock the laser at
resonance $\Delta\og\equiv 0$ with a bandwidth of about $10
\unit{kHz}$. Calibration of the spectra may be performed as
described in the previous section.

While this approach obviously allows reducing the complexity of the
experiment, this arrangement proved more sensitive to slow
temperature drifts in the polarization mode dispersion of the fibres
employed, due to the large ratio of signal and LO powers, the
magnitudes of which are only defined by the polarization state of
the light in the fibre taper region. Improved stability may be
obtained by reducing fibre length to its minimum of ca.\ $0.5
\unit{m}$. For reasons of flexibility and convenience, the actual
fibre length totaled to several meters in our experiment.
Nonetheless, sensitivities of $10^{-18}\unit{m}/\sqrt{\mathrm{Hz}}$
are achieved in toroids using this method \cite{Schliesser2008}. The
intrinsic polarization selectivity of WGM renders the introduction
of an additional polarizer, mandatory in the original implementation
\cite{Hansch1980}, obsolete. In an earlier experiment with a
Fabry-Perot cavity \cite{Hahtela2004}, the losses associated with an
intracavity polarization element limited the finesse, and therefore
the attained sensitivity to $\sim
10^{-14}\unit{m}/\sqrt{\mathrm{Hz}}$.

\section{Observation and analysis of quantum and thermal noises}
\label{s:results}

\begin{figure}[htb]
\centering
\includegraphics[width=.8\linewidth]{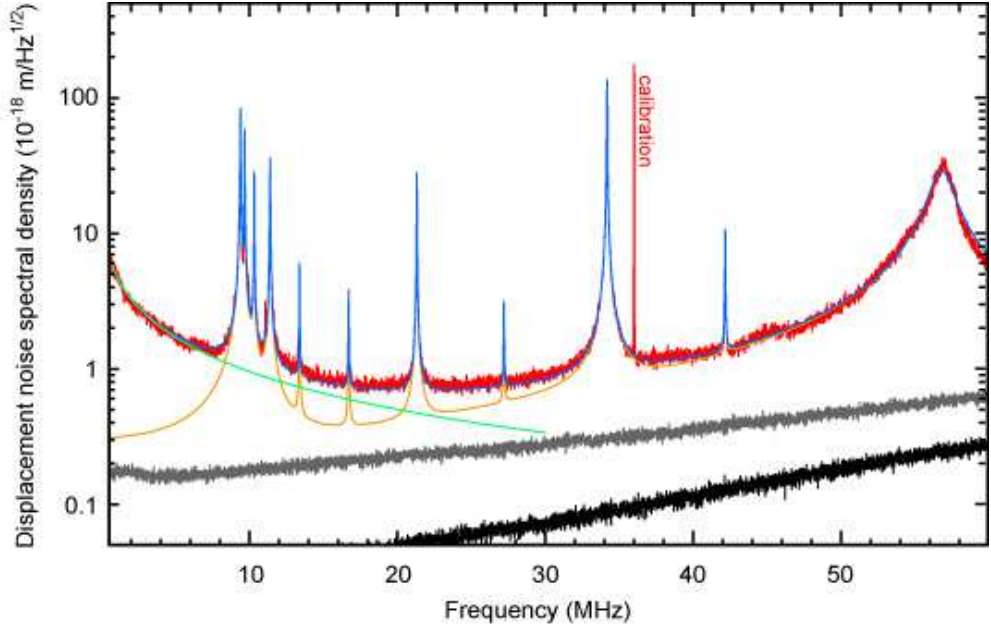}
 \caption{Broadband spectrum of the equivalent displacement noise in a silica toroidal cavity.
 Red, measured trace with laser coupled to a cavity resonance. The peak at 36 \unit{MHz}
 is due to phase modulation for the purpose of absolute displacement calibration.
 Gray, measured shot noise with taper retracted from the cavity.
 The frequency dependence arises from the calculated reduced sensitivity
 at frequencies beyond cavity cutoff. Black, measured electronic detector noise. Orange and green lines are models for mechanical noise and thermorefractive noise,
 respectively, and the blue trace is a sum of the two models and the shot-noise background. }
 \label{f:broadbandcutoff}
\end{figure}

In this section, we present characteristic results obtained with
silica microtoroids of typical major radii between $25 \unit{\mu m}$
and $50 \unit{\mu m}$. \Fref{f:broadbandcutoff} shows an example of
a broadband measurement using homodyne detection. If the taper is
retracted from the proximity of the cavity, quantum shot noise
exceeds the electronic noise in the receiver. It was verified that
the photocurrent noise $\sqrt{S_I}$ scales with the square root of
the total power as expected for shot noise. While this noise is
spectrally flat, the equivalent displacement noise exhibits a
calculated $\sqrt{1+\Omega^2/(\kappa/2)^2}$ frequency dependence
beyond the cavity cutoff at $\kappa/2\approx2\pi \cdot 17
\unit{MHz}$.

When the laser is coupled and locked to a WGM resonance, a
substantially different spectrum is observed
(\Fref{f:broadbandcutoff}). Its equivalent displacement noise is
calibrated in absolute terms using an {\it a priori} known phase
modulation at $36 \unit{MHz}$, and taking the cavity cutoff into
account. The equivalent displacement noise of the cavity exceeds the
shot noise at all frequencies for a high enough power in the signal
beam, leading to a background level equivalent to a displacement
noise of $\sqrt{S_x}\sim10^{-18} \unit{m}/\sqrt{\unit{Hz}}$. The
superimposed sparse spectrum of peaks fits the sum of several
Lorentzians which arise from the thermal noise of several mechanical
modes, $\sum_i |\chi_i(\Og)|^2 S_{F,i}^\mathrm{th}(\Og)$. In the
following, we discuss the features of the spectrum in more detail.

\subsection{Thermorefractive noise}

The broadband, low-frequency background noise is attributed to
thermorefractive noise, the fluctuations in refractive index induced
by the fluctuations $\langle \Delta T^2 \rangle =k_\mathrm{B}
T^2/\rho c_\mathrm{p} V$ of temperature on a microscopic volume $V$
\cite{Landau1980}. Here $k_\mathrm{B}$ denotes Boltzmann's constant,
$T$ temperature, $\rho$ density and $c_\mathrm{p}$ specific heat
capacity. This leads to fluctuations of the resonance frequency via
both the dependence of the refractive index $n$ on temperature, and
the thermal expansion of the material. At room temperature however,
the coefficient of thermal expansion $\alpha$ is more than twenty
times smaller than $dn/dT$, so the analysis can be restricted to the
resonance frequency fluctuations induced by thermo\emph{refractive}
fluctuations.

Introducing a fluctuating thermal source field in the heat diffusion
equation similar to a Langevin approach \cite{Braginsky2000} it is
possible to derive the spectrum of refractive index fluctuations
sampled by a WGM in a silica microsphere \cite{Gorodetsky2004}. For
high frequencies, an approximate analytic expression, neglecting
also the boundary conditions for thermal waves, can be obtained. The
result
\begin{equation}
  S_{\delta n/n}(\Omega)\approx\frac{k_\mathrm{B} T^2 D}{\pi^{5/2} n^2 \rho c_\mathrm{p}
  R}
  \frac{2}{\sqrt{d^2-b^2}}\left(\frac{d n}{d T}\right)^2 \int_0^{+\infty} \frac{q^2 e^{-\frac{q^2 b^2}{2}}}{D^2 q^4+\Og^2}
  \frac{dq}{2\pi} \label{e:trn}
\end{equation}
was found in good agreement with the experimental data obtained on a
silica microspheres \cite{Gorodetsky2004} between $100 \unit{Hz}$
and $100\unit{kHz}$. Here $R$ is the cavity radius, $d$ and $b$ are
transverse mode dimensions and $D$ is the thermal diffusivity  of
silica. For comparison with the toroid measurement calibrated as
effective radial displacement, $R \sqrt{S_{\delta n/n}(\Omega)}$ has
to be evaluated. Inserting the material parameters of fused silica
and the radius of the employed toroid into the model \eref{e:trn},
the data between $100 \unit{kHz}$ and $20\unit{MHz}$ can be
quantitatively reproduced if no parameters except $b$ and the
absolute magnitude are adjusted by factors of order 2
(\Fref{f:broadbandcutoff}). These corrections are justified
considering the approximations made in the derivation, and potential
differences in surface effects in spheres and toroids. It is
interesting to calculate the resulting experimental root-mean-square
fluctuations of the cavity's refractive index
$\sqrt{\int_{-\infty}^{+\infty} S_{\delta n/n}(\Og) d\Og}$ to be of
order $10^{-10}$, as it constitutes a detection limit for resonance
frequency shifts induced by molecules in the evanescent field
\cite{Schroter2008}.

For the purposes of cavity quantum optomechanics, thermorefractive
fluctuations constitute a background noise, at room temperature
rolling off to a level of $\sim 10^{-19} \unit{m}/\sqrt{\unit{Hz}}$
at $\Omega/2\pi>50\unit{MHz}$, where the high-quality
radial-breathing modes typically reside. Practically, such
experiments are going to be performed in a cryogenic environment,
leading to significant changes in the material properties.  A level
of $R \sqrt{S_{\delta n/n}(\Omega)}\sim 10^{-20}
\unit{m}/\sqrt{\mathrm{Hz}}$ at $T\sim 1 \unit{K}$ may be estimated,
assuming a reduction of $dn/dT$ to $\lesssim8\cdot 10^{-6}
\unit{K}^{-1}$ as indicated by recent measurements \cite{Park2007}.
While other sources of noise, such as thermoelastic and photothermal
noises \cite{Braginsky1999, Matsko2007} are not expected to exceed
this value, thorough experimental characterization at cryogenic
temperatures is necessary.

Such a study is also an important pre-study for experiments aiming
at the demonstration of generating broadband intracavity Kerr
squeezing. While above-threshold parametric oscillations have been
observed in these cavities \cite{Kippenberg2004a, DelHaye2007}, the
room temperature experiments reported here indicate that the
thermorefractive noise exceeds the quantum noise of the light in the
cavity, as evidenced by the homodyne measurement of the cavity
output. It therefore has to be suppressed to achieve squeezing of
the cavity field.

\subsection{Mechanical modes}

\begin{figure}[tbp]
\centering
\includegraphics[width=.9\linewidth]{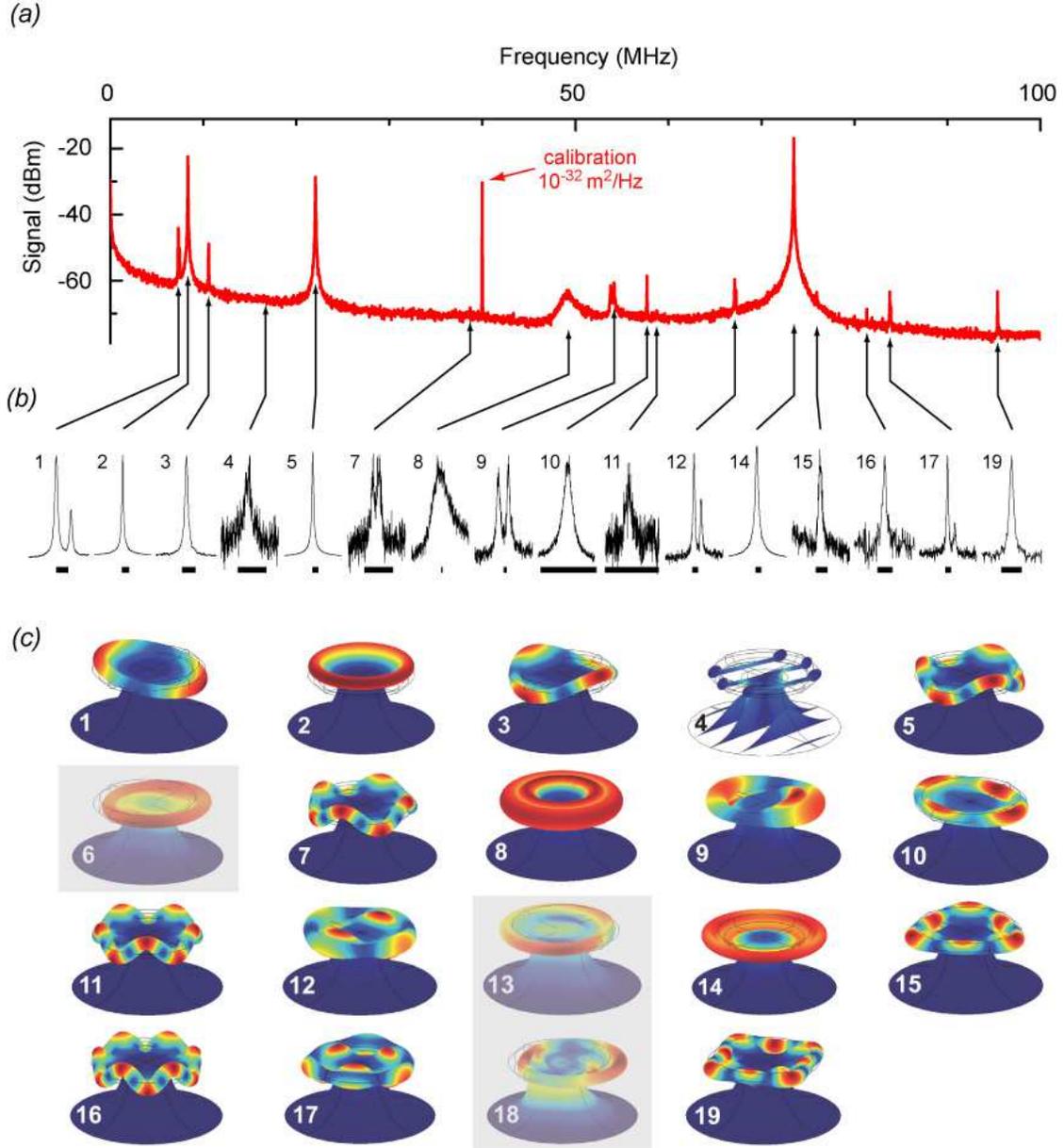}
 \caption{Mechanical modes in a silica toroidal cavity. (a) Displacement noise spectrum obtained by polarization
 spectroscopy. The calibration peak at $40 \unit{MHz}$ corresponds to a displacement spectral density of
 $1.0\cdot 10^{-32} \unit{m^2}/\unit{Hz}$.
 (b) Frequency zoom on 16 Lorentzian peaks identified in the spectrum (linear scale).
 The bar below each spectrum indicates
 a $100\unit{kHz}$ frequency span. The expected two-fold degeneracy of some of the modes is lifted
 (mode 1, 7, 9, 12, 17)  which is attributed to eccentricity.  The panels are enumerated to
 facilitate comparison with (c) showing the spatial displacement patterns
 (highly exaggerated, also indicated in the color code) for the mechanical modes, as obtained from finite element modeling. The
 identification of the modes is made via frequency comparison (see \fref{f:compare}). The pattern of mode
 4 is illustrated showing the displacement of initially parallel slices in order to depict its mainly torsional motion.
 Modes 6,13 and 18 involve mainly motion of the silicon pillar and have not been observed experimentally.}
 \label{f:modesMeas}
\end{figure}

\begin{figure}[htbp]
\centering
\includegraphics[width=.5\linewidth]{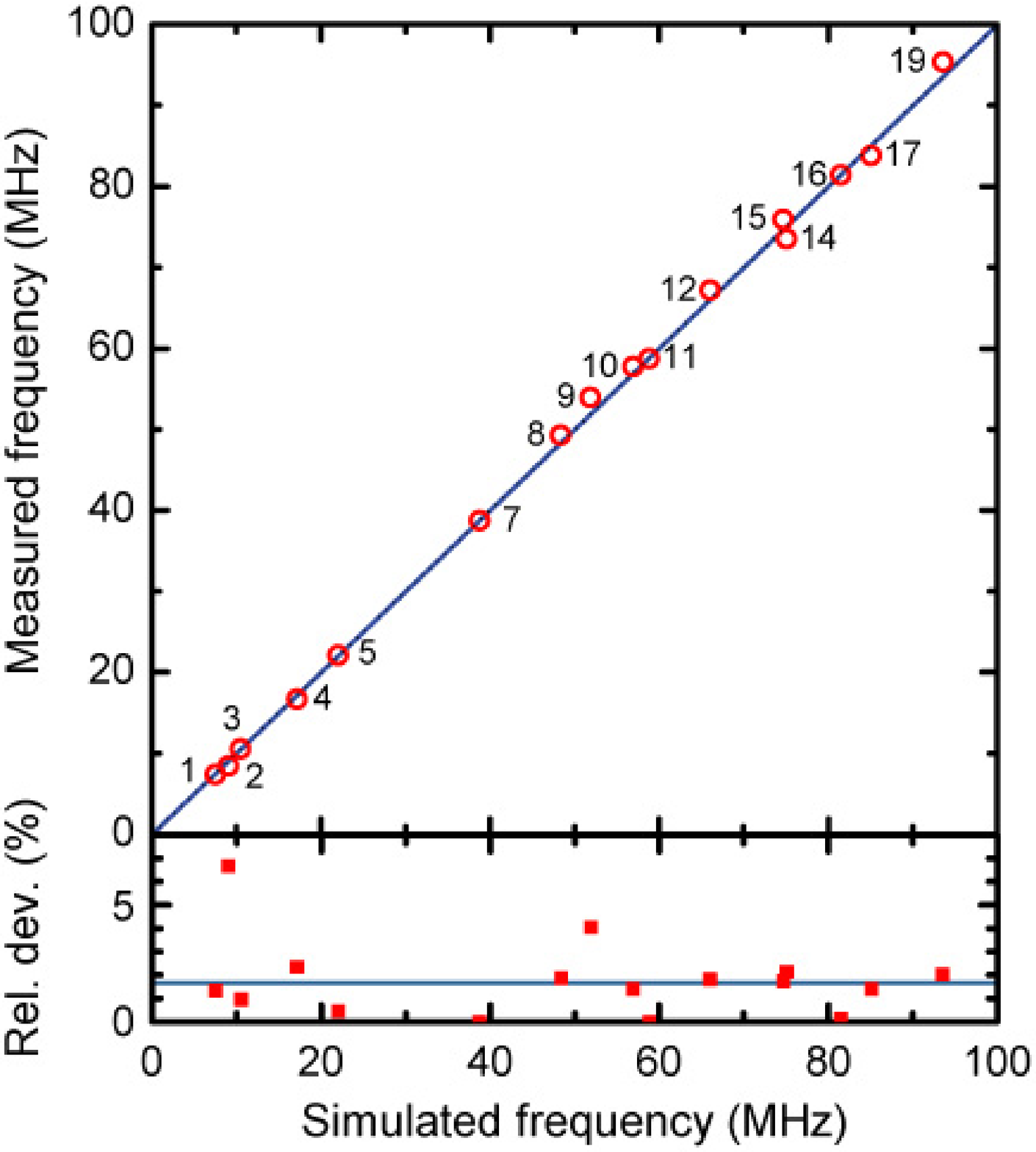}
 \caption{Simulated against measured frequencies of mechanical modes
 (open circles) of monolithic silica microcavities ranging from
$0-100\, \mathrm{MHz}$. The measured frequencies are in excellent
agreement with the values obtained by FEM simulation, showing a
relative deviation of on average less than 2\% (full squares). The
corresponding mode patterns are given in \fref{f:modesMeas}.}
\label{f:compare}
\end{figure}

The most prominent features in the noise spectrum are the mechanical
modes of the microtoroids. To date, the variety of mechanical modes
in silica microcavities (toroids and spheres) has been investigated only to a limited extent%
\cite{Kippenberg2007, Schliesser2008}, in most cases by driving
modes with particularly strong optomechanical coupling into the
parametric oscillatory instability
\cite{Kippenberg2005,Rokhsari2005, Carmon2007, Ma2007, Park2007a}.
Using a Fabry-Perot cavity, however, it was shown that the Brownian
motion of a wealth of intrinsic mechanical modes of a cylindrical
mirror can be studied \cite{Briant2003b}. The high sensitivity
methods presented in the previous section enable monitoring the
Brownian motion of around twenty different mechanical modes in
silica microtoroids over a frequency range spanning from below $1
\unit{MHz}$ to above $100 \unit{MHz}$. Figure \ref{f:modesMeas}
shows a noise spectrum obtained by H\"ansch-Couillaud spectroscopy
revealing a total of 16 mechanical modes. In order to indentify the
observed peaks with the appropriate mechanical mode patterns, a 3D
finite element model (FEM) of the microtoroid is employed and
implemented\footnote[1]{The commercial software Comsol Multiphysics
was employed.}. Extracting the geometry parameters using an optical
microscope (accuracy $\pm 5\%$) and matching observed and simulated
mechanical frequencies, the FEM simulation allows identifying all
observed peaks in the spectrum. Thus all 16 observed modes were
assigned to the corresponding simulated mode patterns as depicted in
\fref{f:modesMeas}. \Fref{f:compare} gives an overview of simulated
frequencies and the frequencies deduced from the spectrum shown in
\fref{f:modesMeas} revealing excellent agreement. The relative
frequency deviation between measurement and simulation is on average
less than 2\%. Moreover, almost all simulated modes are observed
experimentally. Only three out of 19 modes (number 6, 13, 18) cannot
be observed which may be due to low mechanical Q factors ($<10$).

Due to its composite structure comprising several geometric objects
microtoroids exhibit a diverse set of eigenmodes. In order to
characterize the noise spectra, understanding the complex mode
structure of microtoroids is in particular important as all modes
contribute to a background noise floor. Indeed, various
mode-families can be identified in which the motion of the silica
disk, the silica torus, and the silicon pillar partially decouples.
The mode showing the strongest optomechanical coupling is the radial
breathing mode (mode 14 in \fref{f:modesMeas}), and most previous
work has focused on this mode
\cite{Kippenberg2005,Schliesser2006,Schliesser2008}. An equivalent
mode has been studied in a microdisk structure \cite{Clark2005}
where it was termed radial contour mode. In contrast, the
optomechanical coupling of the torsional mode (mode number 4) where
the silica disk shows an in-plane rotation vanishes to first order.
Interestingly, this torsional mode can nevertheless be observed
experimentally.

\begin{figure}[!h]
    \centering
        \includegraphics[width=.5\linewidth]{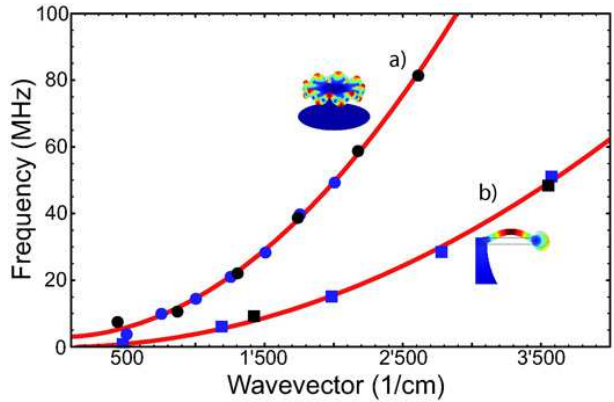}
        \caption{Measured frequencies of crown-modes (full circles) and flexural ``cantilever''
        modes (full squares) of two different samples (black, blue)
         versus their wavevector $2 \pi/\lambda$. Insets illustrate typical displacement patters.
         The frequency dispersion of both mode families allows an accurate quadratic
         fit (a, b) even for data from different samples
          (major radii here: $23\,\mathrm{\mu m}$ and $45 \, \mathrm{\mu m}$).
           This confirms that indeed both mode families can be regarded as
           ring and cantilever modes respectively.}
        \label{fig:Dispersionall}
\end{figure}

One particular mode family that can be identified are the radially
symmetric flexural modes (modes 2, 8) in which the motion of the
free standing part of the silica disk resembles the modes of a
cantilever. The fundamental frequencies of a cantilever of length
$L$ can in general be expressed as $\Omega_i/2\pi=C \cdot
\sqrt{k_i}/2\pi$, where $C$ is a material constant and the $k_i$ are
given by the solutions of \cite{Cleland2003}
\begin{equation}
\mathrm{cos}(k_i\cdot L) \cdot \mathrm{cosh}(k_i \cdot L) + 1 = 0.
\end{equation}
Figure \ref{fig:Dispersionall} shows the measured frequencies of the
two lowest order flexural modes (modes 2, 8) and, in addition, the
first five flexural modes of a different sample plotted as a
function of the wavevector $k_i$, where the free standing part of
the silica disk is taken as equivalent cantilever length $L$
($13.2\, \mathrm{\mu m}$ and $39.6\,\mathrm{\mu m}$ respectively).
Both sets of data allow an accurate single quadratic fit of the
fundamental radially symmetric modes. Thus, the latter can indeed be
regarded as cantilever modes following a uniform quadratic
dispersion even for microtoroids of different sizes. In particular,
the quadratic dispersion rules out the presence of radial tensile
stress within the silica disk as this would imply a linear
dispersion relation.

Another obvious mode family which can be distinguished is
characterized by sinusoidal oscillations of the torus itself (modes
1, 3, 5, 7, 11, 16). The dispersion diagram of these modes, which we
refer to as crown-modes, is depicted in \fref{fig:Dispersionall} for
two different samples. The respective wavelength $\lambda$ is given
by twice the distance between two adjacent nodes of each mode. The
frequencies $\Omega/2 \pi$ of the crown modes observed in
microtoroids of different circumference and torus diameter
($2\pi\cdot23\,\mathrm{\mu m}$/$5.3\,\mathrm{\mu m}$ and
$2\pi\cdot45\,\mathrm{\mu m}$/$5.7\,\mathrm{\mu m}$) allow a
\textit{simultaneous} quadratic fit with the frequencies $\Omega_i
/2 \pi$ following a quadratic dependence on the wave vector
$k_i=2\pi/\lambda_i$. This uniform dependence shows that in this
mode family the silica torus, despite its attachment to the silica
disk, behaves effectively like an independent element. The observed
quadratic dispersion relation rules out the presence of tensile
stress within the torus in radial direction as this would lead to
linear dispersion characteristic of a vibrating string. Since the
microtoroids undergo a reflow process \cite{Armani2003} indeed all
potentially present stresses should get relaxed during this
fabrication step. As such, the observed quadratic dispersion
characteristic for a rigid ring structure confirms this expectation.

\begin{figure}[tbp]
    \centering
        \includegraphics[width=.5 \linewidth]{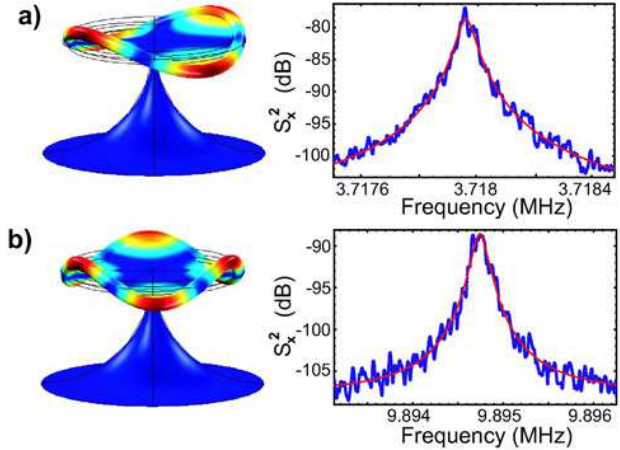}
        \caption{Spectra and simulated mode patterns of the second and
        third order crown modes. They exhibit very high mechanical
        quality factors of 58,000 and 51,000, which is attributed to
        low clamping losses \cite{Anetsberger2008}.}
        \label{fig:CrownModes}
\end{figure}

The study of clamping losses in the radial breathing mode and the
optimization of its mechanical Q factors \cite{Anetsberger2008} also
lead to the observation of very high Q crown modes. For these
measurements, the cavities are operated in low-pressure environment
($p<1 \unit{mbar}$) in order to reduce viscous damping, which
limited the Q's of the previously discussed modes to values
$\lesssim 3000$. For such a measurement, \fref{fig:CrownModes} shows
the second and third order crown modes with Q factors exceeding
50'000 at frequencies below $10\, \mathrm{MHz}$. These high Q
factors are attributed to low clamping losses which are studied in
detail in \cite{Anetsberger2008}.

\section{Conclusions}

In conclusion, we have shown that high-finesse whispering gallery
modes are extraordinarily well suited as transducers for
micromechanical motion. Sensitivities on the order of $10^{-19}
\unit{m}/\sqrt{\mathrm{Hz}}$ are achieved, on par with the best
reported values \cite{Arcizet2006, Caniard2007a}. The small
dimensions of the WGM resonators allow in addition extending the
measurement bandwidth by more than an order of magnitude compared to
earlier work \cite{Arcizet2006, Caniard2007a}. This enlarged range
conicides with the resonance frequencies of the mechanical modes of
interest present in the device. For example, the radial breathing
mode (RBM), particularly amenable to optomechanical effects due to
its small effective mass ($\meff\sim 10\unit{ng}$), typically
exhibits a resonance frequency around $50 \unit{MHz}$. At this
frequency, average thermal phonon occupancies below unity are
achieved at temperatures $\sim 2 \unit{mK}$. Approaching such
temperatures appears feasible using a combination of conventional
cryogenics and resolved-sideband cooling \cite{Schliesser2008}. It
is interesting that the expected spectrum of even the zero-point
fluctuations of this mode peaks at a value of
$\sqrt{S_{x}^{\mathrm{ZPF}}(\Omega_{\mathrm{m}})}=\sqrt{\hbar
Q/\meff\Om^{2}}\sim 10^{-19} \unit{m}/\sqrt{\mathrm{Hz}}$, assuming
an effective $Q=100$ after cryogenic and laser cooling. Such signal
levels may be detected on top of the background of the thermal
noises studied here, providing a route towards experimental tests of
the theory of quantum measurements on mesoscopic objects. Finally,
we note that the advantageous properties of WGM resonators may also
be exploited for motion transduction of a mechanical oscillator
external to the cavity, for example, by bringing it into the near
field of the whispering gallery mode.

\section*{Acknowledgements}
The authors acknowledge discussions with T.~W.~H{\"a}nsch and J.\
Kotthaus. T.~J.~K. acknowledges support through an Independent Max
Planck Junior Research Group Grant, a Marie Curie Excellence Grant
(JRG-UHQ) and the DFG-funded Nanosystems Initiative Munich (NIM).
The authors gratefully acknowledge J.~Kotthaus for access to
clean-room facilities for microfabrication.

%\section*{References}

\bibliographystyle{unsrt}
\bibliography{C:/Dokume\string~1/Albert/Eigene\string~1/Literature/microcavities}

\end{document}